\documentclass{article}
\usepackage{amsmath,amsfonts}
\usepackage{algorithmic}
\usepackage{algorithm}
\usepackage{array}
\usepackage[caption=false,font=normalsize,labelfont=sf,textfont=sf]{subfig}
\usepackage{textcomp}
\usepackage{stfloats}
\usepackage{url}
\usepackage{verbatim}
\usepackage{graphicx}
\usepackage{cite}
\usepackage[tight]{units}
\def \be {\begin{equation}}
\def \ee {\end{equation}}
\begin{document}
\title{Electron quantum optics with beam splitters and waveguides in Dirac Matter}
\author{Derek Michael Forrester$^{\ast}$\textit{$^{a}$}and Fedor Vasilievich Kusmartsev\textit{$^{b, c}$}}
\maketitle
\footnotetext{\textit{$^{a}$~QinetiQ, Advanced Services \& Products, Cody Technology Park,Farnborough, Hampshire, GU14 0LX; E-mail: DMForrester@QinetiQ.com}}
\footnotetext{\textit{$^{b}$~College of Art and Science, Khalifa University, Abu Dhabi P.O. Box 127788, UAE; E-mail: fedor.kusmartsev@ku.ac.ae }}
\footnotetext{\textit{$^{c}$~Loughborough University, Physics, Epinal Way, Loughborough, Leicestershire, LE11 3TU}} 
\begin{abstract}
An electron behaves as both a particle and a wave. On account of this it can be controlled in a similar way to a photon and electronic devices can be designed in analogy to those based on light when there is minimal excitation of the underlying Fermi sea. Here splitting of the electron wavefunction is explored for systems supporting Dirac type physics, with a focus on graphene but being equally applicable to electronic states in topological insulators, liquid helium, and other systems described relativistically. Electron beam-splitters and superfocusers are analysed along with propagation through nanoribbons, demonstrating that the waveform, system geometry, and energies all need to balance to maximise the probability density and hence lifetime of the flying electron. These findings form the basis for novel quantum electron optics. 
\end{abstract}
\section{Introduction}
Advances in electronics have led to the possibility of confining or maneuvering electrons for applications ranging from quantum information processing \cite{Acciai2019} to  single-electron quantum optics and nanoelectronics \cite{Bocquillon2012}.  Single-electron emitters are highly desirable quantum technologies to create devices analogous to those developed for quantum optics, leading to the new field known as electron quantum optics. The propagation of single electrons can be controlled by beam splitters, interferometers, and waveguides, e.g. focusing electron flow with graphene Veselago lenses \cite{Cheianov2007,hills2017current}. A Lorentzian voltage pulse, or other energy input of similar pulse shape, applied to a 2D material has been shown experimentally to emit one electron without affecting the local background Fermi sea \cite{Dubois2013}. This creates an electronic soliton without relying purely on electron confinement \cite{Keeling2006}. In the current work an investigation of propagating electrons with dissipative soliton behaviour is carried out for electrons pulsed through various waveguides. The splitting of the wavefunction occurs in two arms of a beamsplitter, which could form part of a Hong-Ou-Mandel interferometer \cite{HOM1987,Burset2019}.  The similar two arms geometry has been used in the phenomena known as Josephson fluxon-cloning observed in superconductors \cite{gulevich2006flux,gulevich2006new}. There the superconducting wave function associated with the fluxon has been split and after splitting two identical fluxons  were propagating parallel in two identical arms.

Graphene is an ideal material for exploring such single electron physics. It has amazing properties such as huge electron mobility and ballistic transport. These properties are associated with the electron massless Dirac spectrum as shown in Ref.\cite{kusmartsev1985semimetallic}.  Due to these properties the electron wave packet can propagate over long distances without much scattering.  Moreover such electron wavepackets can be detected with the present microscopy tools. Indeed, recently high amplitude peaks in the wave function of quantum dots in bilayer graphene have been observed by using a scanning tunneling microscope \cite{Velasco2020}. This enabled observation of the wave behaviour of the electron wavepacket in bilayer graphene flakes. Similarly, single layer graphene on hexagonal boron nitride has been examined using STM techniques and  spatial control of charge doping created for on-demand p-n junctions \cite{Velasco2016}. Other work has shown that electron transport can be controlled over femtosecond scales by applied light and a bias voltage, leading to tuning of electron thermalisation pathways in graphene \cite{Ma2016}.

In graphene nanoribbons electrons are expected to act like photons in optical waveguides, without dissipation at the edges, or like fluxons in long Josephson junctions\cite{gulevich2006flux}-\cite{gulevich2019bridging}. Only the presence of lattice defects and impurities may create dissipation at the edges of the material. Here the edges of a nanoribbon  can guide electrons via edge states, which is an effect that is usually happening in a magnetic field (e.g. \cite{liu2015snake}). The situation can be different when an electron is ejected with an energy higher than the Fermi energy. Moving in the bulk of the nanoribbon, however, such an electron wavefunction can spread when the width is large enough, appearing to dissipate as the probability density decreases. In general the  boundaries of the nanoribbon can form a Fabri-Perot resonator for such an excited electron. Thus, such tapered waveguides are examined here in order to explore the electron quantum optical effect of superfocusing. In freestanding pristine graphene, electrons behave as if they are massless (and, therefore have no band gap). It has been shown that the band gap is a function of the width, length, and edge states of a graphene nanoribbon due to confinement of $\pi$ electrons \cite{ribbons2017}. Bottom-up approaches such as on-surface and solution-phase polymerisation for synthesizing nanoribbons have been preferred to methods such as e-beam lithography for precise control over width, length, and edge definition \cite{Yano2020}. Typically top-down approaches, such as lithographic cutting of graphene, tend to produce defects at the edges of the material, whilst bottom-up approaches lead to smooth edges and controllable widths that are determined by the precursor molecules used. The electrical (and magnetic) properties of the nanoribbons depend on the type of edge they have and the chemical structure, with zig-zag edge states favourable for spintronics and armchair edge states tunable through bandgap engineering at room temperature \cite{Liljeroth2015}. The narrowest possible metallic, zero bandgap armchair nanoribbons were synthesised by Liljeroth and colleagues \cite{Liljeroth2015}, being just five carbon atoms wide. These types of nanoribbon fall into the $N=3p+2$ family, where $p$ is an integer and $N$ is the number of carbon atoms. There exist another two families, the $N=3p$ and $N=3p+1$ types, which are semiconductors with wide bandgaps. Recently $N=17$, $3p+2$ family nanoribbons on Au(111) have also been produced using dibromobenzene-based precursor monomers (as well as $N=13$ structures) \cite{Yamaguchi2020}. All compositions of nanoribbons are now achievable using these bottom-up approaches with fusing of neighbouring $N=7$ armchair compositions achieved for $N=14$ and $17$ reported too \cite{Wee2012}. Ballistic transport in $\unit[40]{nm}$ wide epitaxial graphene nanoribbons has also been described \cite{deHeer2014}, demonstrating sheet resistance below $\unit[1]{Ohm/sq}$, and channels of the order of micrometers in length. This is quite remarkable for electron transport over such distances without any dissipation, pushing the potential for electron quantum optics into reality. The self-assembled growth of single layer graphene on the etched trenches in high quality silicon carbide wafers produces the waveguides for ballistic transport, i.e. the electron moving without scattering\cite{ben2020raman}. In the next section an investigation of graphene nanoribbons is performed and a possible mechanism for controlled transport explored.
\begin{figure}[!t]
\begin{center}
\includegraphics[width=9.5cm,keepaspectratio]{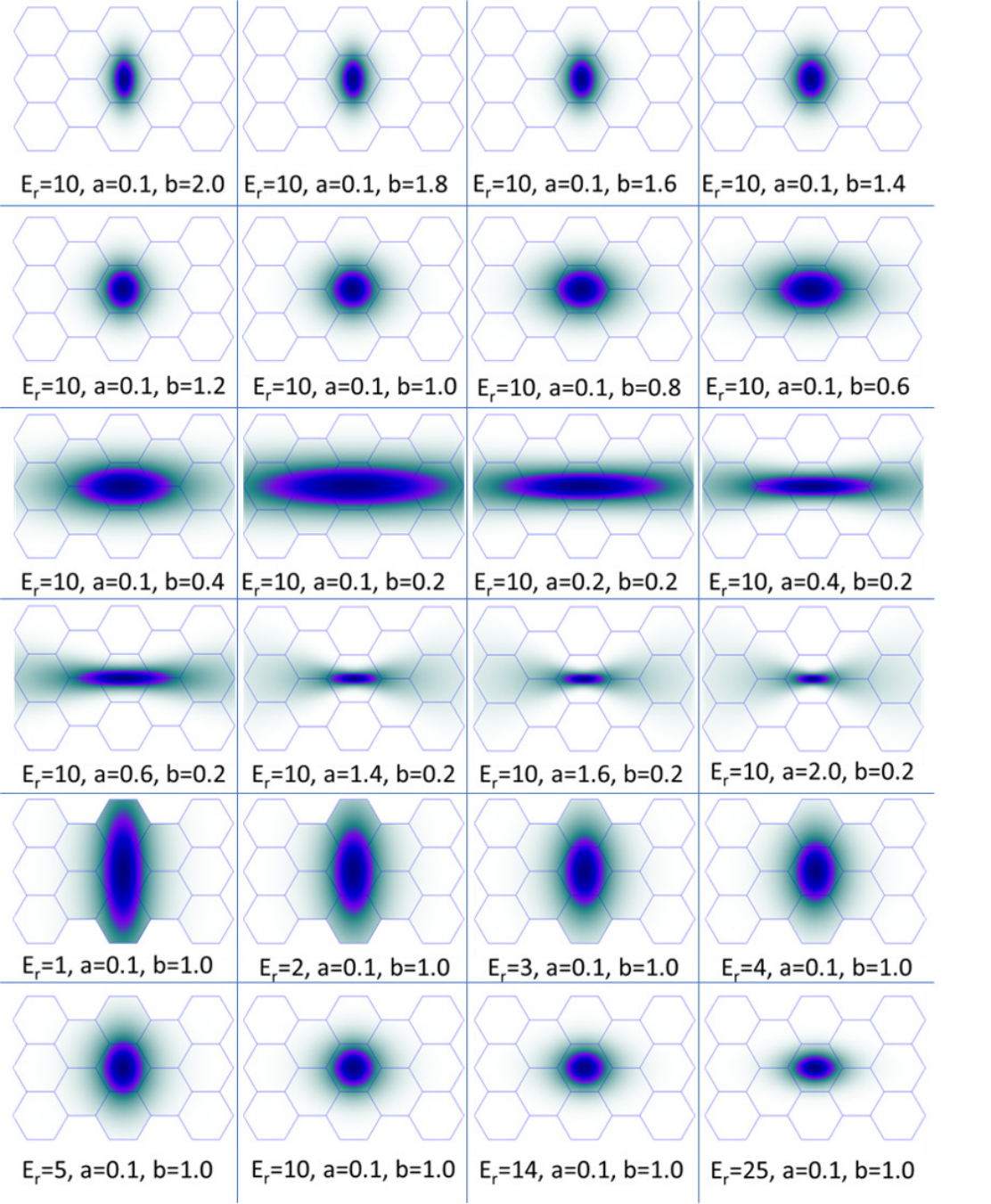}
\end{center}
\caption{The form of the electronic wavepacket at initiation, $t=0$, as a function of the effective mass parameter $E_r$ and the shape parameters $a$ and $b$ of the Gaussian beam (normalised by characteristic length $a_0$). The actual value of the electron energy is $E=E_r m_{gr}^* v_{Fgr} v_{Fnr}$, where $m_{gr}^*$ is the effective mass of sheet graphene (subscript gr), taken to be $0.012M_0$ \cite{effmass}, where $m_0$ is $\unit[9.109\times10^{-31}]{kg}$  . The Fermi velocity is taken to be, $v_{Fgr}=v_{Fnr}$ with a value of $\unit[10^6]{m/s}$, where subscript $nr$ stands for nanoribbon. The form is shown through the  probability density with central peak equal to one (darkest blue).} 
\label{FigureA}   
\end{figure} 
\section{Graphene nanoribbons}
As discussed above there have been significant efforts made to produce graphene nanoribbons (GNRs) through chemical synthesis. The two main types of GNR are ones with an armchair or zig-zag edge. There exist other configurations based on side branches of carbon atoms and alterations to the structure that develop during growth. A very promising approach to producing GNRs is the patterning of silicon carbide before epitaxial growth of graphene. This avenue of research has high promise for industrial scaling for very high performance graphene electron optics. To understand the flow of electrons in narrow graphene waveguides GNR structures are now explored. The system can be described in terms of the Dirac equations \cite{Nanoscale2015, RSCAdvances2015,NPGR2013}. As such, we solve the system of equations,
\be 
\frac{\hbar }{i}\partial_t \left(\begin{array}{c} \Psi_1 \\ \Psi_2 \end{array}\right) =
H\left(\begin{array}{c} \Psi_1 \\ \Psi_2 \end{array}\right)\label{eq:dirac}
\ee
where,
\be
H=\nu_F\left (\begin{array} {cc} 0 & -i\hbar \partial_x-\hbar\partial_y \\ -i\hbar \partial_x+\hbar\partial_y & 0 \end{array} \right)\label{eq:ham}.
\ee
In the above, the single electron motion is represented by the wavefunction $\Psi_1$ as a second electron wave $\Psi_2$ maneuvers to fill any void created by the former and to produce a complete Fermi sea. The electron is kicked up from the Fermi sea through the correct voltage or photo-excitation giving an initial Lorentzian waveform moving with Fermi velocity $\nu_F$ and electron energy $E_r$,  propagating through the system to take on a Gaussian shape \cite{Nanoscale2015,RSCAdvances2015}. We now write the dimensionless form of Eq. (\ref{eq:dirac}),
\be
-i\partial_\tau \left(\begin{array}{c} \Psi_1 \\ \Psi_2 \end{array}\right) =
\bar{H}\left(\begin{array}{c} \Psi_1 \\ \Psi_2 \end{array}\right)\label{eq:diracd}
\ee
and
\be
\bar{H}=\left (\begin{array} {cc} 0 & -i \partial_{\bar{x}}-\partial_{\bar{y}} \\ -i \partial_{\bar{x}}+\partial_{\bar{y}} & 0 \end{array} \right)\label{eq:hamd}.
\ee 
Note that in these differential equations (\ref{eq:dirac}), $\hbar$ stands on both sides of equations and, therefore, is cancelled here.
Time is written in dimensionless units as, $\tau=\nu_F t/a_0$, and the propagation distances given in 2-D space as $\bar{x}=x/a_0$ and $\bar{y}=y/a_0$, where $a_0$ is a characteristic length, e.g. in graphene, a lattice constant of $\unit[0.246]{nm}$. These systems are typically two-dimensional electron gases where the electrons are free to move in two-dimensions. In Figure\ref{FigureA}, the shape of the Gaussian beam is shown at the point of propagation through a ribbon for various values of effective mass and shape factors. Its initial form is strongly dependent on the effective mass of the electronic soliton in the nanoribbon.
\begin{figure}[!t]
\begin{center}
\includegraphics[width=9.5cm,keepaspectratio]{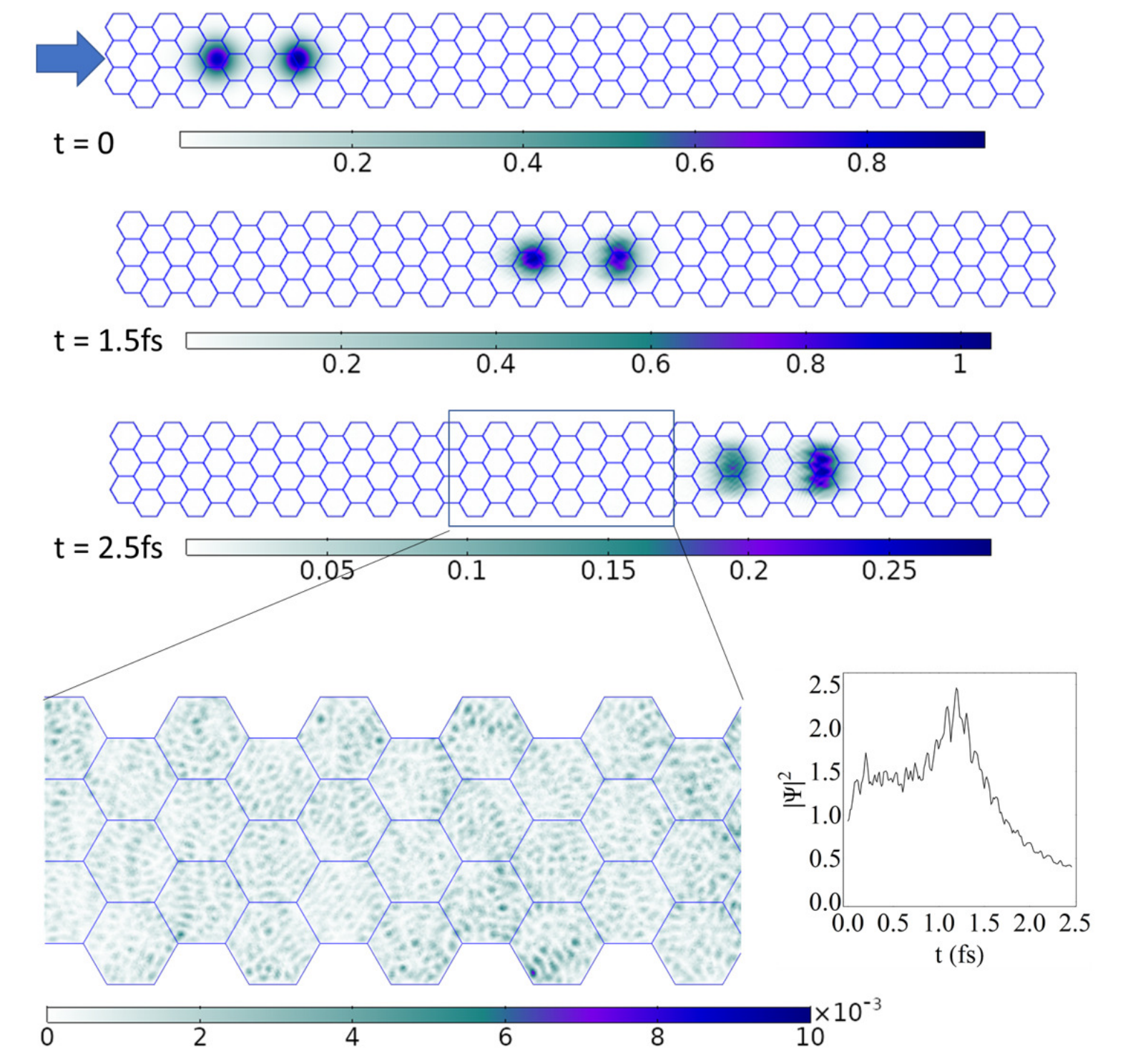}
\end{center}
\caption{A pulse of two electrons travelling through an $N=8$ armchair graphene nanoribbon. The top three images are snapshots of the propagating wavepacket at measurement times $t=0$,  $\unit[1.5]{fs}$, and $\unit[2.5]{fs}$}. 
\label{FigureB}   
\end{figure}
In Figure \ref{FigureB} a GNR with width $N=8$ demonstrates the spread of the wavefunction even in a very narrow channel for a pair of propagating electrons. Although the focus of this work is on single electron propagation, pulses of electrons can be incorporated too. In this case the width of each electron wavepacket is less than the width of the channel and is still propagating mainly through the bulk of the ribbon more than interacting with the edge. The maximum probability density of the two propagating solitons becomes diminished along the channel over the time of $\unit[2.5]{fs}$. Figure \ref{FigureB} illustrates the wave-like nature of the electron as well as the particle like behaviour - in the bottom left image the interference of the wavefunction within the channel is shown on a reduced scale $\left|\Psi\right|^2<10^{-2}$. This clearly illustrates that in this particular $N=8$ nanoribbon rapid dissipation will emerge.This is in agreement with experimental observation where nanoribbon edge effects produce departure from ideal metallic behaviour but now with the added complexity of the soliton. The initial characteristic frequency of the electron is defined based on de Broglie's definition so that at velocities less than the speed of light the electron energy is dominated by its rest energy $E_0=m_0c^2$. Electrons in graphene have an effective mass, however, of $m_{gr}^*=\unit[0.012]{m_0}$ \cite{effmass} indicating an increased ease of transmission through potential barriers or steps. The electron kinetic energy in graphene, near the Dirac points, is equal to the effective mass times the Fermi velocity squared (with the group velocity, $v_g$, and phase velocity $v_p$, equal to each another and the Fermi velocity, $v_F^2=v_g v_p$). Thus, $E=m_g^* v_F^2$, where $m^*$ is the effective mass. The effective mass in armchair nanoribbons, such as that of Figure \ref{FigureB}, is also a function of the width of the ribbon, with all families such as $3p+1$ and $3p+2$ types exhibiting tunable bandgaps. For the single flying electron it is assumed that the Fermi velocity in the nanoribbons is equal to that in large width monolayers,  $\unit[10^6]{m/s}$. The energy of the emitted electron is given in dimensionless units as $E_r=m_{nr}^* v_{Fnr}/m_{gr}^* v_{Fgr}$, where the subscripts $nr$ and $gr$ refer to the graphene nanoribbon and graphene continuous layer, respectively. Thus, after formation of the electron wavepacket the wavefunction is described by,       
\be
\Psi_1=\frac{cosh \big[b\left(\tau-\bar{x}-L\right)\big]\exp\left[{-\left(E_r\left(\tau-\bar{x}-L\right)-\frac{E_r a \bar{y}^2}{2\left(1+i~a\bar{x}\right)}\right)}\right]}{\sqrt {\left(1+i~a~\bar{x} \right)} }\label{eq:phi1},
\ee
and
\be
\Psi_2=\frac{\left(1+a \bar{y}\right)}{\left(1+i~a\bar{x}\right)} \Psi_1\label{eq:phi2}.
\ee
\begin{figure}[!t]
\begin{center}
\includegraphics[width=8.5cm,keepaspectratio]{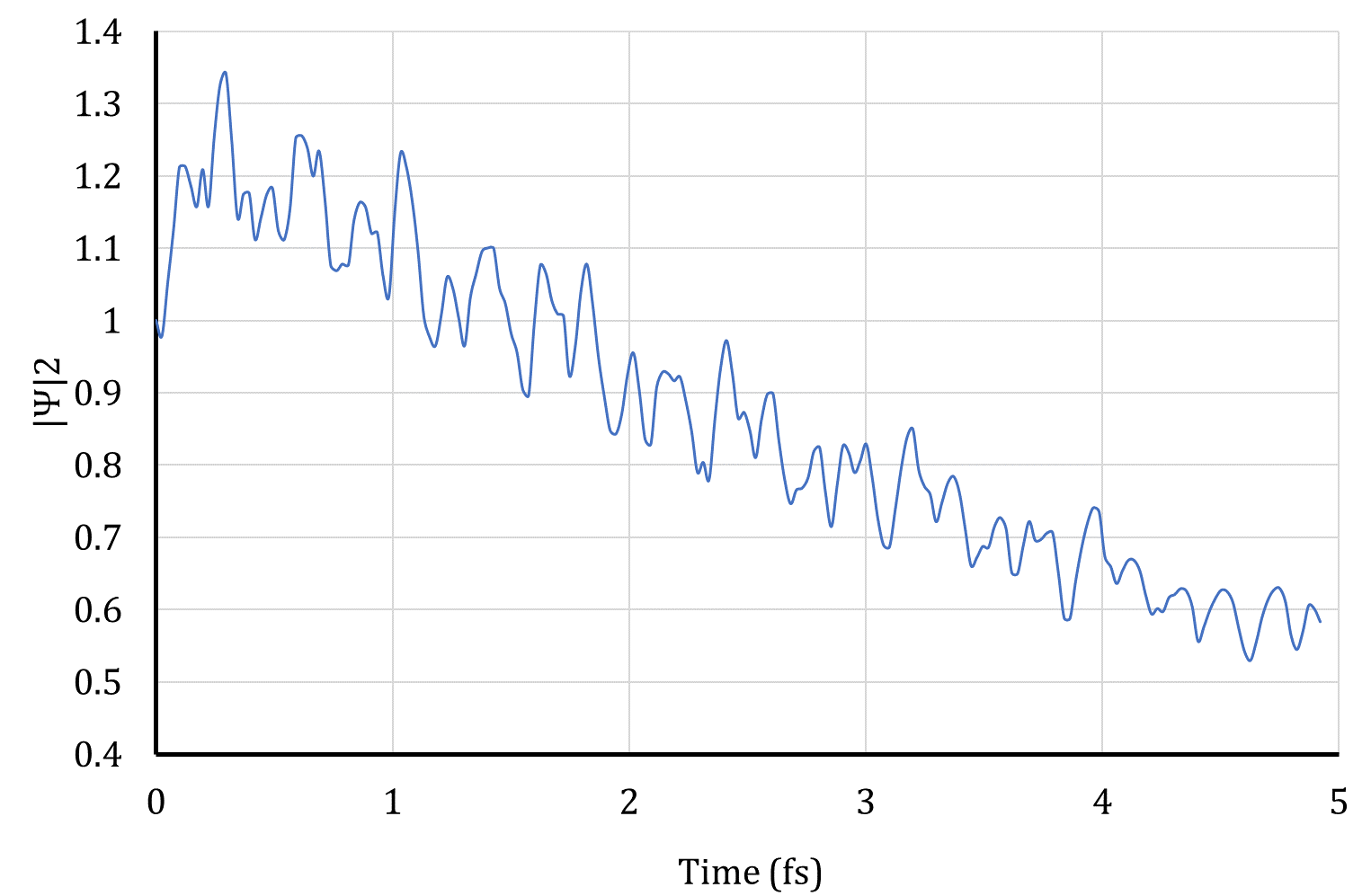}
\end{center}
\caption{An electron wavepacket travels through an $N=7$ nanoribbon The probability density as a function of time is shown to $t=5 fs$.}
\label{FigureC}   
\end{figure}
The shape of the wavepacket is defined by ${a}$ and ${b}$ at $t=0$, given after an initial arrival time. In Fig. \ref{FigureC}, the minimally excited flying electron is defined with parameters $a=0.1$, $b=1.6$, and $E_r=25$ (corresponding to $1.7eV$ and $m_{nr}^*=0.3 m_0$, which is close to the values found experimentally for $N=7$ \cite{Ruffieux2012}).
The use of  Gaussian beam methods for localised wave  problems is considered  to be very powerful and illustrative of physical processes.  They have been found to be very useful for descriptions of the localised excited states  in many electronic systems such as quantum dots and wires and in particular in graphene resonators and nanoribbons \cite{zalipaev2008high}.It was also beneficial for describing levitons and anti-levitons in our previous paper \cite{Nanoscale2015}. The benefit of the usage of  Gaussian wave packets is that the  Gaussian wave function has  exactly a Gaussian similar shape as wave function of electrons localised on artificial atoms such as quantum dots. Precisely speaking, it is an exact solution for the wave function for electrons trapped by parabolic potential wells. Here, we may also use another approach.  In the alternative approach, electrons  are even more localised than in a Gaussian wave packet: in the tight binding approximation electrons are tightly bound with atoms,  generating hopping between them. Definitely that form of electron splitting can be well described  in such a model. However, in all existing experiments Levitons are created by injection of  electrons which have a Lorentz shape at the their initiation. It is  spread on many atoms. After injection that transform collapses (almost instantaneously) into Gaussian form - the Gaussian wave packet -  as in excited states  in many electronic systems and in particular in graphene resonators and nano-ribbons \cite{ferrari2015science}\cite{meunier2016physical}. Note that similar transformation also happens for photons, with the creation of photon Gaussian wave packets that are associated with single photons \cite{deng2022observing}. For this reason we use the Gaussian  wave packet to describe the flying electrons injected in our nanodevices.  When the band is narrow,  the approach will be equivalent to that where electrons will be localised on atoms, and then we will use a tight binding approximation,  i.e., a tight-binding model. However, in narrow bands the levitons have not been observed so far. For this reason we have not used another approach  and  although we used localised  states of electrons as Levitons for the Gaussian wave packets, we believe that our results about electron beam splitting is universal and does not depend on the model.

In equations (\ref{eq:phi1}) -(\ref{eq:phi2}), $L$ characterises the initial position of the beam, which is arbitrary. In our simulations we use hard-wall boundaries. The reason for that is that there is a large work function from graphene to vacuum larger than a few electron  volts\cite{yu2009tuning}, and due to this work function on the boundaries the formation of effective hard walls occurs, from which electrons are reflected. 
\begin{figure}[!t]
\begin{center}
\includegraphics[width=9.5cm,keepaspectratio]{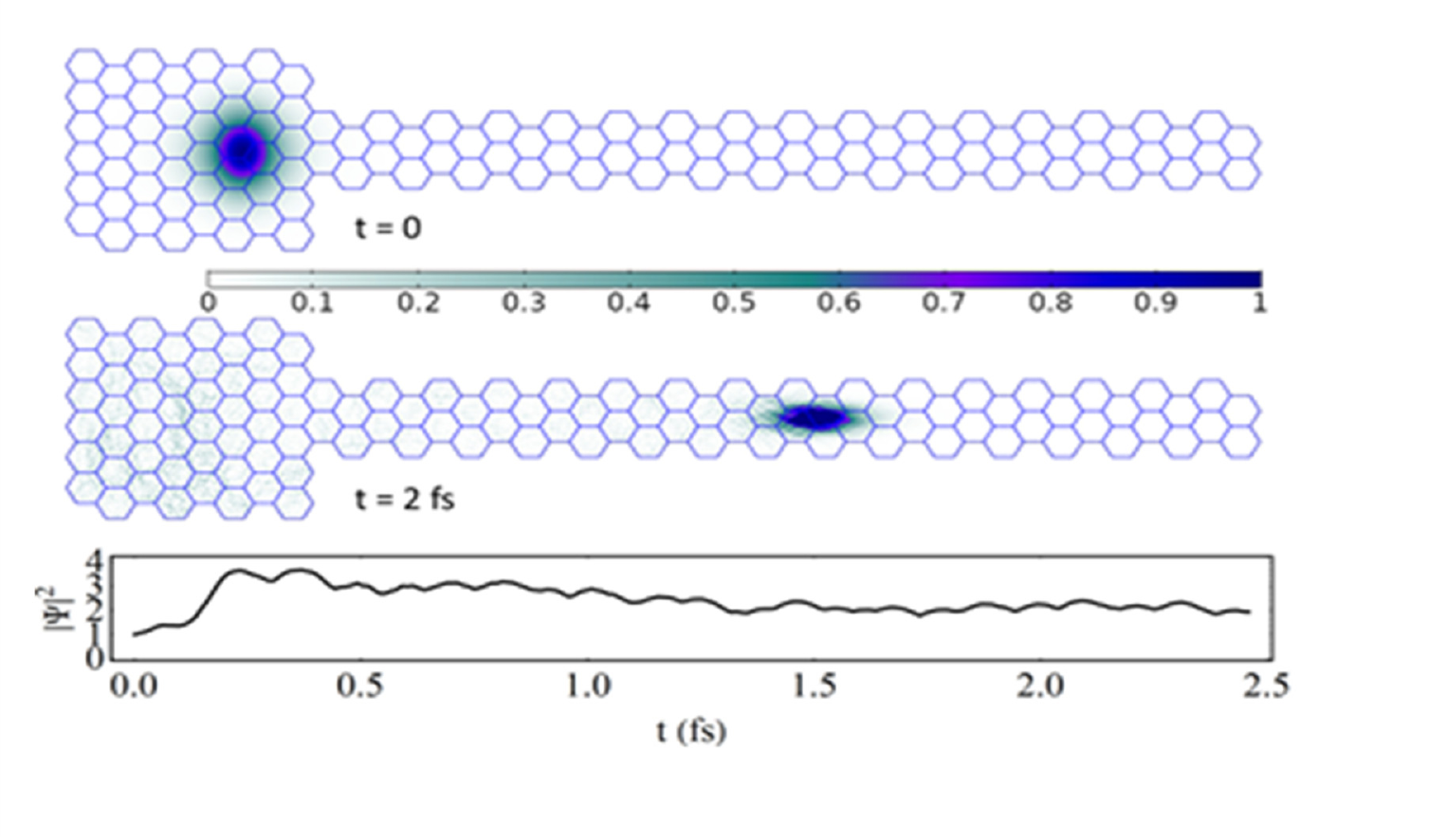}
\end{center}
\caption{An electron wavepacket formed in an $N=14$ nanoribbon and travelling through an $N=6$ armchair graphene nanoribbon. The bottom plot shows the probability density as a function of time, with the above images being shapshots of the evolution occurring at $t=0$ and $t=\unit[2]{fs}$}
\label{FigureD}   
\end{figure}
\begin{figure}[!t]
\begin{center}
\includegraphics[width=8.5cm,keepaspectratio]{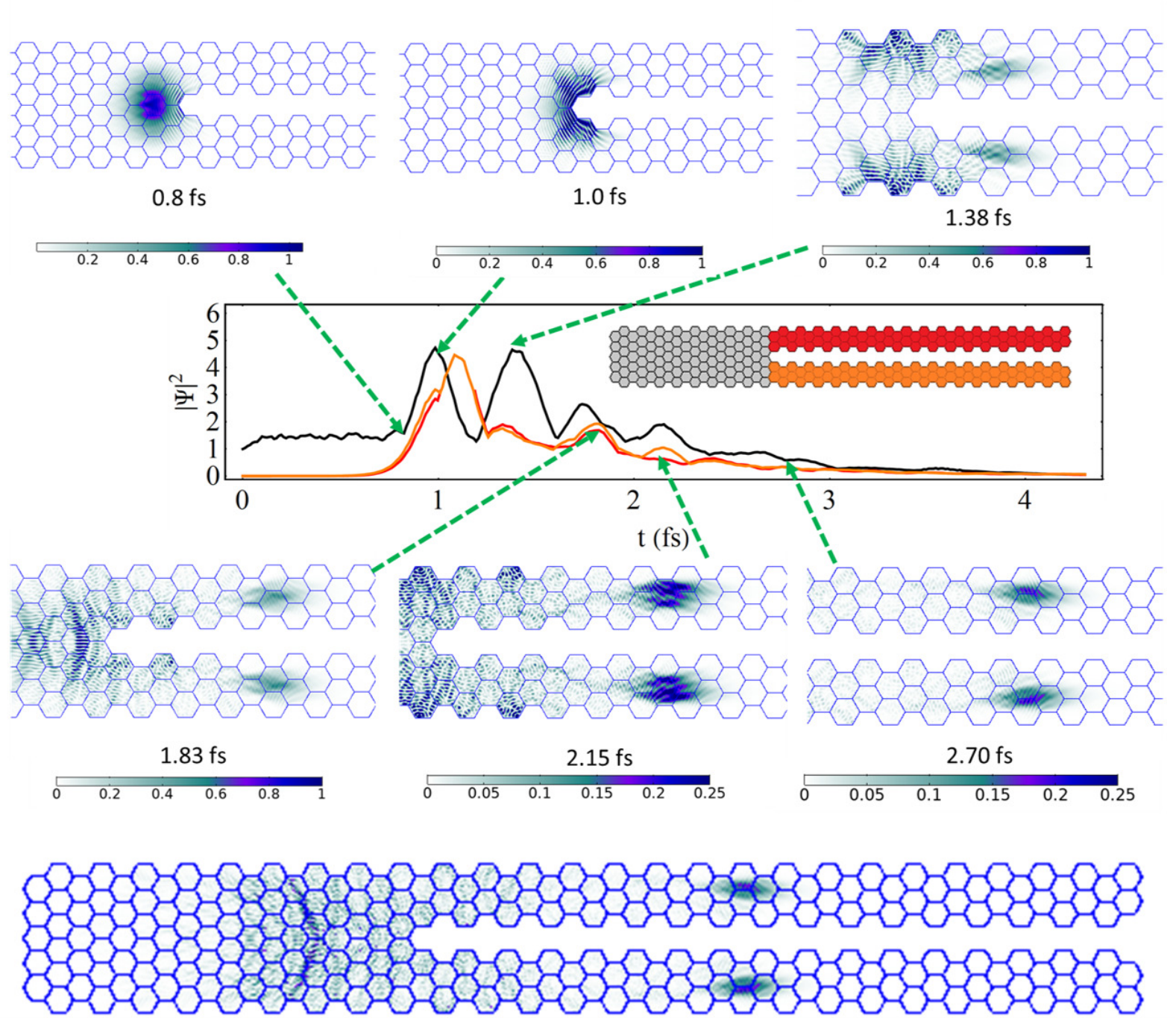}
\end{center}
\caption{An electron wavepacket travelling in two $N=6$ armchair graphene nanoribbon channels split from an $N=13$ area.The maximum probability density is  plotted against time (black line, whole system; red line, just in the top channel; orange line, just in the bottom channel). The time evolution is shown at times $\unit[0.8]{fs}$ to $\unit[2.7]{fs}$, with the bottom figure also at $\unit[2.7]{fs}$.
The propagating Gaussian wave packet  describes the moving single particle localised in space. Its position is identified by the “maximum probability density”. Strictly speaking, in quantum mechanics we can not determine precisely the electron location due to the Heisenberg  uncertainty relation. However, due the large speed of flying electrons- the effective speed of “light” or the Fermi velocity the electrons - momenta are  larger and therefore the uncertainty in the electron position is shrinking.   Therefore, the effect of the electron beam splitting  can be and is most clearly presented with the use of this concept. Another point is that in the Gaussian packet the  probability density away from the maximum drops to zero very fast. Therefore, the position of the maximum is also important to indicate where the electron is localised despite its quantum mechanical uncertainty. Due to this reason the concept of the “maximum probability density” is useful, in particular, for the description of the electron beam splitter.}
\label{FigureE}   
\end{figure} 
Next an electron formed at a source in a nanoribbon of width $N=14$, connected to a $N=6$ nanoribbon, is examined (Fig. \ref{FigureD}).
One possibility for driving the existence of the electronic soliton (a leviton) is to initiate it by use of a quantum point contact approach.  After the initial formation the levitons (and/or anti-levitons) travel as near-solitons and dissipate as they propagate - hence the spreading of the wavefunction seen in the figures by the reduction in the probability density with time. In our model the actual time is a function of the Fermi velocity and characteristic length of the system. One can see that confinement effects enhance the amplitude of the maximal probability density (e.g. superfocusing) or that splitting of the wavefunction occurs in the nano-dimensioned channels. This creates the analogous behaviour to that observed in quantum optics with beam splitters and confinement effects.

\begin{figure}[b!]
\begin{center}
\includegraphics[width=8.5cm,keepaspectratio]{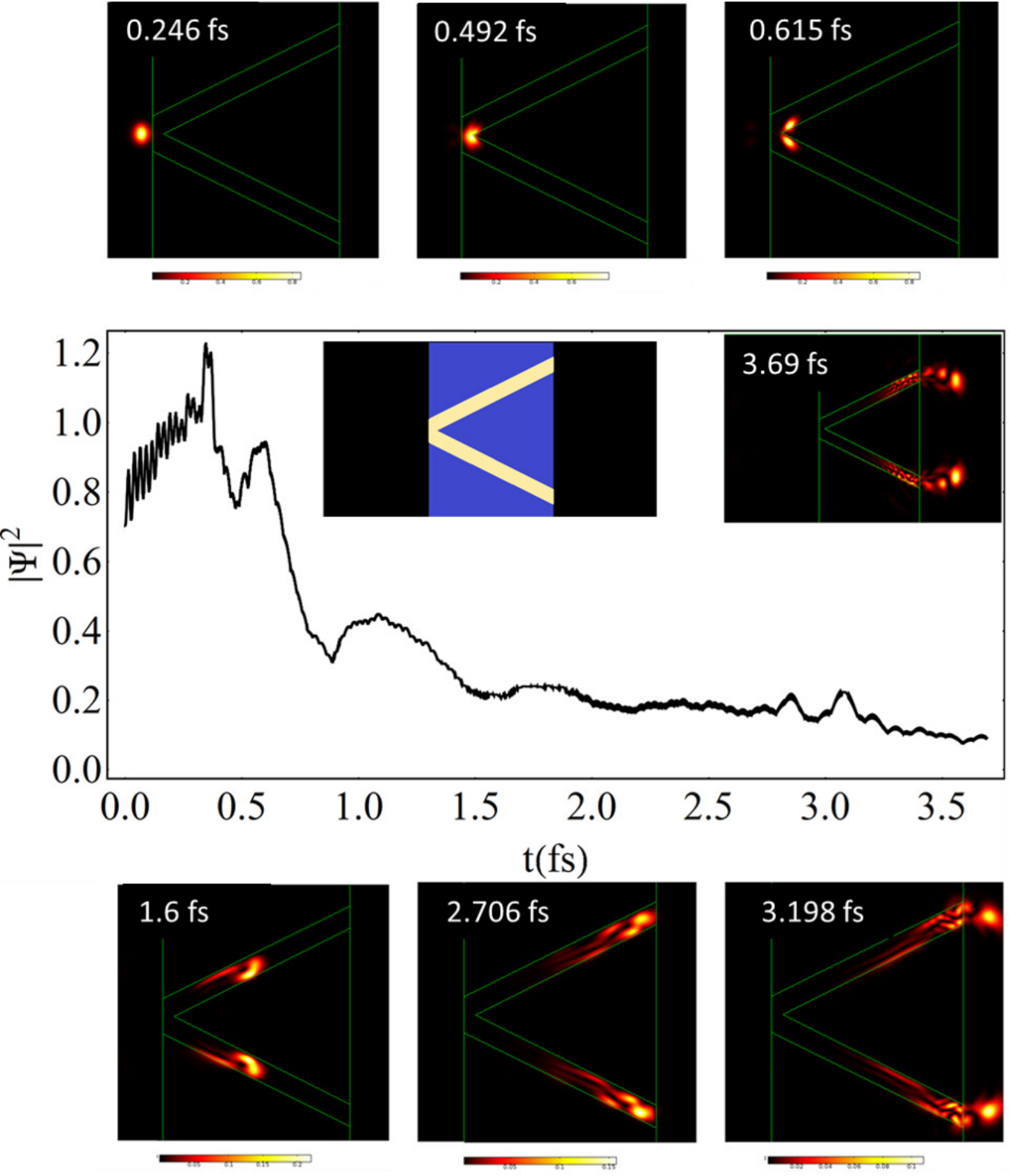}
\end{center}
\caption{The flying single electron propagates towards a V shaped waveguide. The evolution in the two waveguide channels up to $\unit[3.69]{fs}$ is shown through the probability density.}
\label{FigureF}   
\end{figure}

The dynamical evolution of the probability density (shown in Figure \ref{FigureD} is explored for $a=0.025$, $b=1$, and $E_r=25$. In this case the electron wavepacket is squeezed into the $N=6$ channel and the probability density remains greater than one as it travels over $\unit[2.5]{fs}$ (shown in \ref{FigureD}) and beyond.The narrow $3p$ type nanoribbon was chosen as it is to illustrate low dissipation in the soliton system, as is demonstrated though the high probability density. As it propagates there is a small oscillation in its trajectory, marked by fluctuations in $\left|\Psi\right|^2$. A nanoribbon structure similar to Figure \ref{FigureD} is shown in Figure \ref{FigureE}, except with two branches composed of armchair nanoribbons (rather than one) of the  $N=6$ variety. The propagating electron splits to occupy both branches but also has a significant amount of backscatter as it meets the branch intersection. In Figure \ref{FigureE}, the overall probability density (black lines) is plotted alongside the probability density inside the top (red) and bottom (orange) branches. The splitting can be obtained through physically different branches of the nanoribbon as in Figure \ref{FigureE} or through the introduction of regions of different potential energy.
One approach for raising a potential barrier is to create diverging pathways from local deformations in graphene, with interference generated by scattering centres and electrons injected through an STM tip \cite{Juan2011}. Or, for example, the local Fermi velocity can be tuned through modification of the effective dielectric constant of the graphene and substrate \cite{Hwang2012}. Here, a look at a V-shaped electron optical beam splitter is undertaken, with an example shown in Figure \ref{FigureF}.

The V has a blunt interface which marks the transition into a region of higher potential from the region on the left. In the inset of the probability density against time plot of Figure \ref{FigureF}, a colour schematic of the waveguide regions is shown; the blue areas have equal energy with the electron (acting as an exclusion zone) and the light yellow areas have energy $U_p=E_r/\alpha$. The parameter $\alpha$ defines the ratio between the electron energy and that of the potential. In Figure \ref{FigureF}, $\alpha=0.67$, with $a=0.25$, $b=2$, and $E_r=10$. The blunt nose of the V is $~3.0a_0$ wide and the branches of the V are $\sqrt{3}a_0$ wide. The electron wavefunction (in the form of the probability density) is shown outside the waveguide in Figure \ref{FigureF} (top left) at $~\unit[0.25]{fs}$, meeting the blunt nose shortly afterwards and entering the waveguide before splitting down the channels. In each channel the fraction of the electron moves forward with an oscillatory motion. In Figure \ref{FigureF} the potential drops to zero again at the end of each branch/channel and the entangled electronic waves are emitted in a spreading arc. Aharonov-Bohm type rings have also been investigated whereby the split components recombine, marked by an elevation in the probability density.
\begin{figure}[!t]
\begin{center}
\includegraphics[width=8.5cm,keepaspectratio]{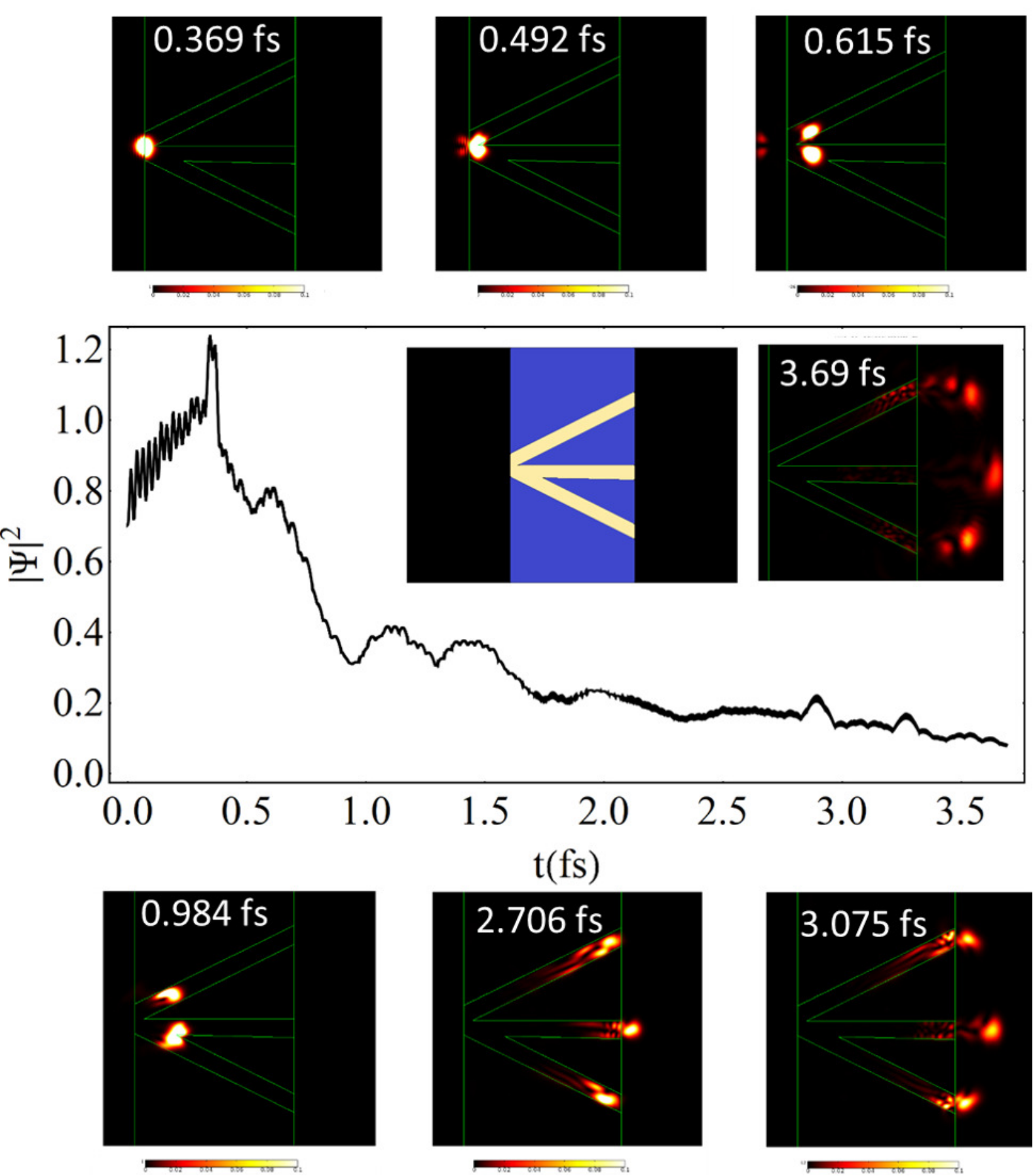}
\end{center}
\caption{A single electron propagates towards the triad shaped waveguide. Shown is the probability density over time, with snapshots of the system probability density at times $\unit[0.369]{fs}$ to $\unit[3.69]{fs}$.}
\label{FigureG}   
\end{figure}
More channels can be incorporated into the system, with an example of three off-set branches and wave propagation shown in Figure \ref{FigureG}. In this example a time delay between the three split electron components of the wavefunction occurs on account of slightly differing channel lengths. The channels have $\alpha=2/3$, $a=0.25$, $b=2$, and $E_r=10$, as in Figure \ref{FigureF}. There are two splittings of the wavefunction at the intersections of the channels at $~\unit[0.49]{fs}$ and $~\unit[0.98]{fs}$. In this case the probability density peaks occur in three regions - channels 1-3 - and form a tripartite entanglement. Further channels could be used to produce multi-partite quantum information protocols. It has been shown that electron propagation is a function of the channel width and that the wavefunction can be split in these systems. In all cases confinement effects lead to an increase in the probability density \cite{RSCAdvances2015} and so the waveform is best targeted at interaction with the edges. 
\begin{figure}[t]
\begin{center}
\includegraphics[width=8.5cm,keepaspectratio]{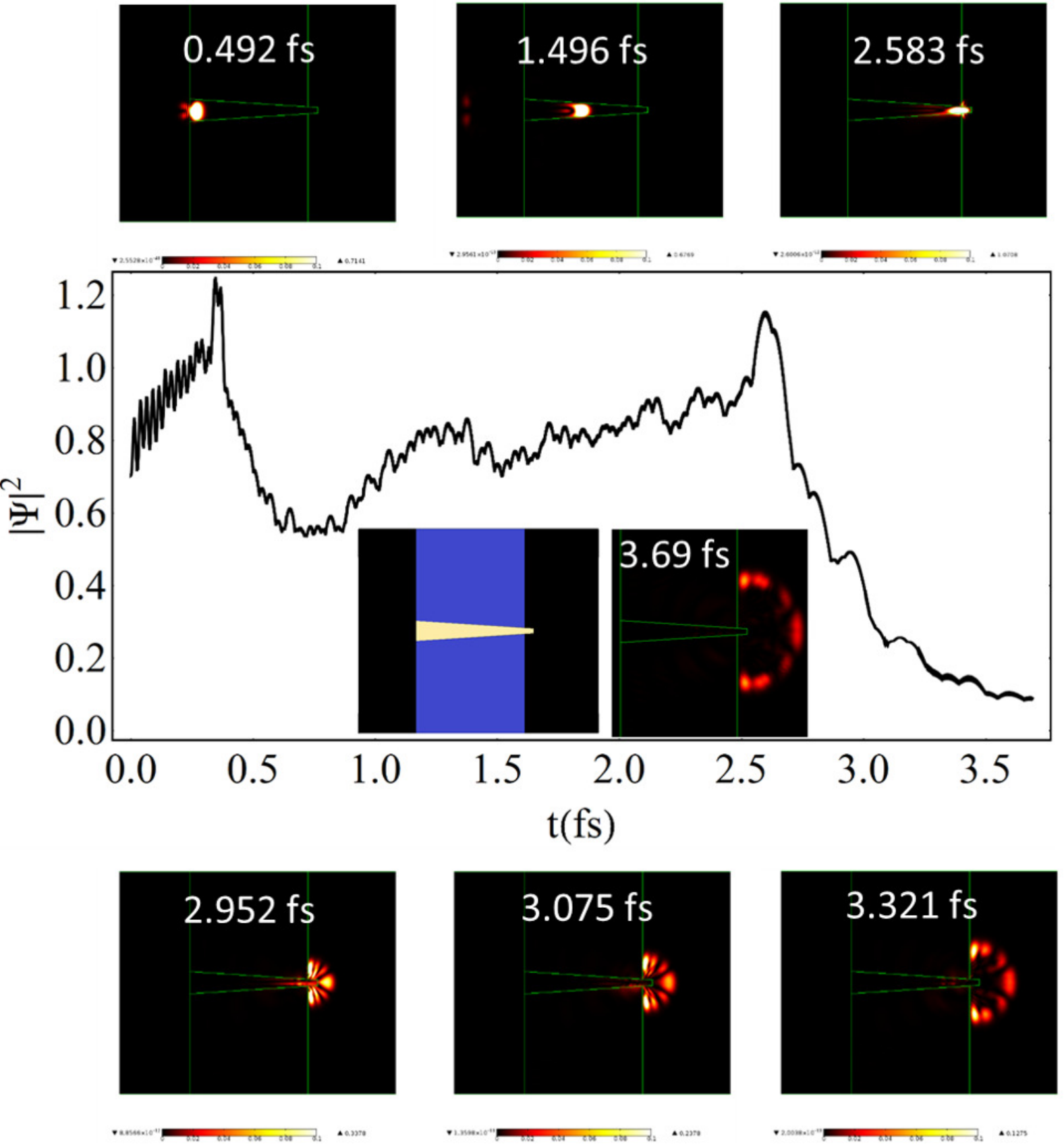}
\end{center}
\caption{The single electron emitted, as a minimal excitation of the Fermi sea, propagates towards and through the the tapered superfocusing ''magic wand'' shaped waveguide. The probability density is plotted up to $\unit[3.7]{fs}$.}
\label{FigureH}   
\end{figure}

In quantum optics superfocusing is used where subwavelength image resolution is required in order to observe nano or micro features in detail \cite{Johnston2007}. The superfocuser can take the form of a tapered structure with a superfocus at its tip. In Figure \ref{FigureH} a superfocusing tapered design for electronics is used to exploit confinement effects and maintain a high probability density with an enhanced electron focus at the end. The length of the superfocuser is $18a_0$, the width at the entrance is $3a_0$ and at the exit $a_0$, with $\alpha=2/3$. The superfocuser is build inside a potential barrier, with $\alpha=1$ (blue region in the inset of the central plot of Figure \ref{FigureH}). Upon emission from the tip, which slightly lies outside the barrier, parts of the wavefunction slide up the barrier wall whilst the remaining $\left|\Psi\right|^2$ is released in an arc. The probability density at $~\unit[2.6]{fs}$ is greater than one. Outside of the barrier the wavefunction is free to spread as there is no confinement potential. 
\begin{figure}[!b]
\begin{center}
\includegraphics[width=8.5cm,keepaspectratio]{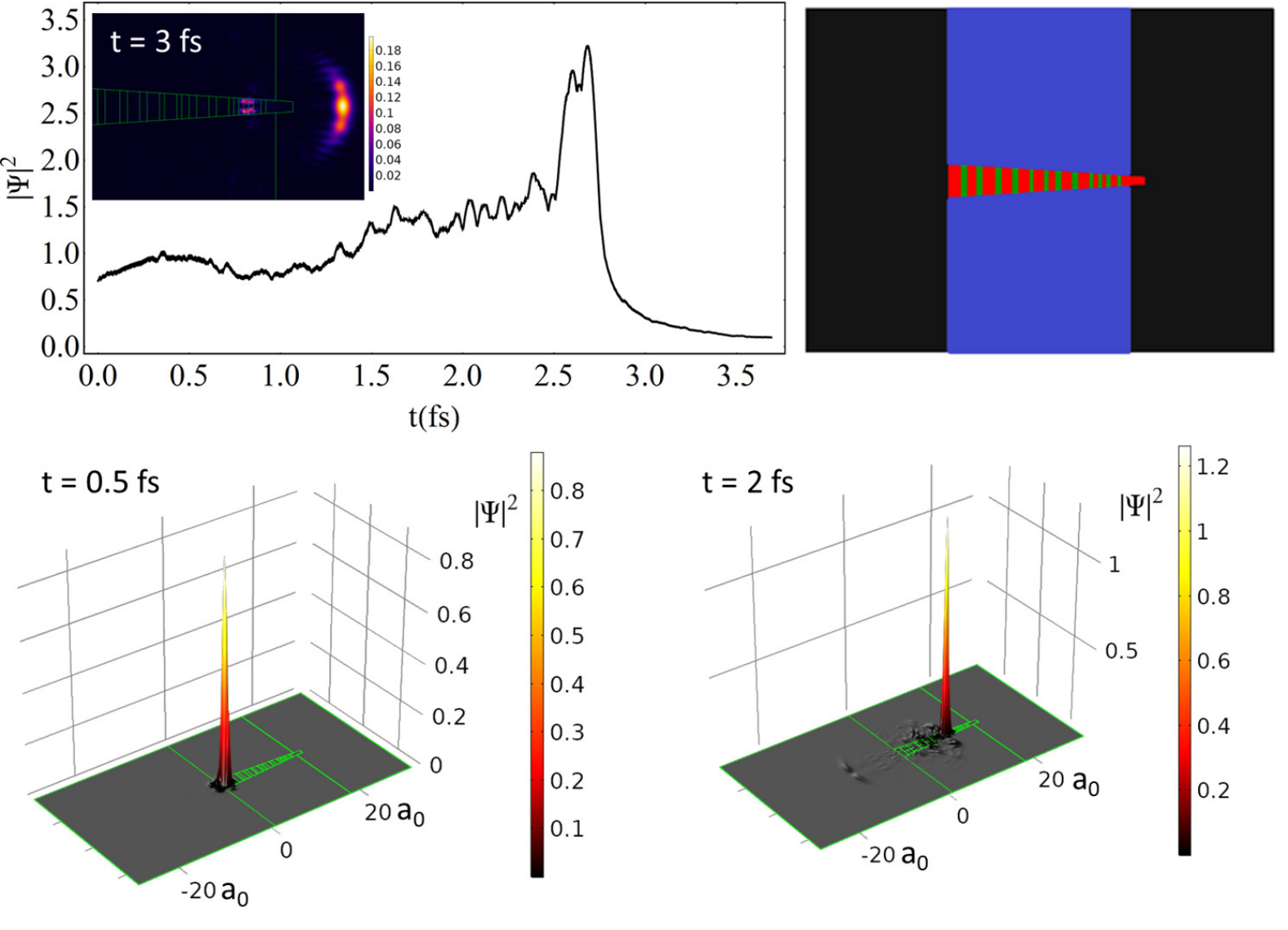}
\end{center}
\caption{Atop the Fermi sea the propagating electron meets with a  tapered superfocusing heterostructure. Examined is the probability density through the superfocuser within a potential step. The blue area has $\alpha=0.35$ and the electron energy level is $E_r=7$. Within the tapered waveguide the red regions have $\alpha=0.47$ and the green areas $U_{p2}=0.35$. The time evolution is shown over $~\unit[3.7]{fs}$.}
\label{FigureI}   
\end{figure}

A small superfocusing heterostructure is now described and shown in Figure \ref{FigureI}. The green and blue areas in Figure \ref{FigureI} have the same potential, which means there is a small leakage through the gaps. However the red areas of the heterostructure with $\alpha=0.47$ completely confine the wavefunction and so by the time the wavefunction has reached the tip the probability density has increased to $3.5$ (the transmission into a potential step at various energy ratios $\alpha$ is described in detail in \cite{Nanoscale2015}). Again the emission takes on the arc pattern seen in the superfocuser in Figure \ref{FigureH}. Thus, combinations of tuned potential steps can be used to boost the signal of the travelling wave. Focusing of electrons can be achieved through negative refraction \cite{Cheianov2007, RSCAdvances2015, Nanoscale2015} Veselago lenses or, as described here, through tapered waveguides, with or without integration of a heterostructure. 
\section{About creation of Majorana Fermions}
In a formal way, the original Gaussian packet has the  wave function $\Psi ( x,y,t)$.
 That corresponds to the probability density to find electron, $|\Psi ( x,y,t)|^2$.   If after beam splitting of the electron wave packet we have two real (Majorana) fermions, then  $\Psi (x) = \Psi_1 (x) +i \Psi_2 (x)$,  where  $i$ is an imaginary unit and for a simplicity of notation we put  $x$ to stands for all space and time variables $(x,y,t)$. Then the final probability density of the two split wave packets will be equal  to $(\Psi_1 (x))^2 + (\Psi_2 (x))^2$,  which is identically equal to $|\Psi (x)|^2$, that is $|\Psi ( x)|^2=\Psi_1 ( x))^2 + (\Psi_2 (x))^2$. Whence  the  electron uniform splitting  may corresponds to the creation two real Majorana Fermions. If instead of the wave function we take the electron field operator $\Psi ( x)^{\dagger} ( \Psi ( x) )$ that corresponds to the creation (or annihilation) of the electron in the space with the  anti-commutation relation.
\be
				\{\Psi ( x)^{\dagger} , \Psi ( y) \}= \delta(x-y)
\ee
 then  its splitting into the two real fermions will correspond to the commutation relation
\be
\{\Psi_1 ( x) , \Psi_1 ( y)\}= \delta(x-y) {\rm ~~and~~ } \{\Psi_2 ( x) , \Psi_2 ( y)\}= \delta(x-y) 
\ee
This corresponds to the creation of the two Majorana fermions. Thus, with the use of this commutation relation  one may indicate that the two new Gaussian packets created after splitting of the original single electron wave packet can correspond to the formation of two real (Majorana) Fermions. Each Majorana Fermion is a half of the original electron. Note, that this necessary condition for the formation of Majorana fermions is satisfied. The question is, will this always happen or not or is this condition of the electron beam splitting into two identical real electrons enough that we will have two Majoranas - this requires more investigations that we will aim to present in future work.
\section{Entangled electron states}
The effect of electron beam splitting described here has a strong potential to create entangled quantum states in electrons, which can have significant implications for developing new technologies based on quantum information. Entanglement is a phenomenon where two or more particles become `quantum correlated', i.e., their quantum states become correlated even when they are separated by large distances.

In particular, the increasing the density in an electron beam can most efficiently lead to the formation of electron entangled quantum states. When two electrons are in parallel nano-ribbons, they can form an entangled state, where they behave like a single unit even when they are separated in different nano-ribbons. This is analogous to two separated Majorana fermions combining into an electron. Such electron splitting is highlighting the unique properties of electrons in quantum nano-systems.

Compared to photon entangled states, which are not confined to nanoscales and are more difficult to integrate into small devices, the entangled electron states are confined to nanoscales and can be integrated into very small nanodevices. This property makes them highly attractive for use in designing quantum devices and new information technologies.

Thus, the huge potential of electron beam splitters to create entangled quantum states in electrons can have significant implications for developing new technologies based on quantum information. Entangled states are a key feature of quantum mechanics, and the unique properties of electron entangled states compared to photon entangled states makes them highly attractive for use in designing quantum devices and specifically nano-devices.

\section{Discussion}
Within this work quantum electron optics in narrow graphene channels, nanoribbons, have been discussed. The splitting of the electron wavefunction into nanoribbon branches has been shown, alongside superfocusing. These discoveries may be important for quantum technologies, such as room temperature quantum computers. Recently, nanoribbon field-effect transistors have been also been demonstrated \cite{FET2017}. New methods in single electron emission and detection are highly sought after for single flying electron devices such as spin and charge qubits. Unlike in photonics, the quantum electron optics are usually hindered by the propagation of the electrons occurring with interaction with the Fermi sea  - thus losing coherence and information over short distances. The unique properties of the minimal excitation soliton mean that it surfs atop the Fermi sea without loss of information, under the right conditions. Approaches to generate single electrons riding on sound waves have also offered new potential for generation of entangled electron pairs over relatively large distances \cite{SAWsurfer}. The surface acoustic wave upon which the flying electron rides is initiated by application of a microwave excitation through exploitation of the piezoelectric properties of the substrate. This is another example of stimulated single electron minimal excitation of the surrounding environment. Indeed there are now a number of methods employable to emit single electrons that surf the Fermi sea with minimal disruption or fermionic interaction. One is the aforementioned surface acoustic wave methods \cite{SAWsurfer} and another is pulse shaping to Lorentzian initial form of the bias voltage to produce levitons \cite{Levitov2006}. Others are the driven mesoscopic capacitor methods in the quantum Hall regime and local gate modulation of a quantum Hall edge state \cite{Dashti2019}. In these specific devices quantum point contacts are effectively used as half silvered mirror analogues and quantum Hall edge channels operate as waveguides \cite{Ferraro2018}. In graphene and other 2D materials, tunable electron sources for quantum electronics are an active area of development \cite{Wilmart2016}. In flying qubits (e.g. single electrons) the information is encoded into the electron trajectory. This has been demonstrated in the current work for splitting of the electron wavepacket into separate branches, which in effect produces fractionalisation of the electron with entanglement of each component (see Figures \ref{FigureF} and \ref{FigureG} for example). It is interesting that for crossed graphene armchair nanoribbons, injected electrons have been shown to split with almost zero back-reflection when the intersection angle of the nanoribbons is $60^o$ - an electron beam-splitter \cite{Brandimarte2017}. Additionally, graphene nanoribbons between superconducting contacts can form a novel form of tunable Josephson junction with a $\pi$-phase shift \cite{Liang2008,aGNRJJ2007}. This type of junction produces a spontaneous current and is ideal for operating as the functional quantum computing element, e.g. for a quantum phase gate \cite{PhaseQubit2016}. The splitting of the wavefunction occurs as the nanoribbon splits into two or more channels, producing several pathways for transmission. In an Aharonov-Bohm type ring the pathways can be made to recombine. In work on superconducting devices, beamsplitters of the superconducting wave function describing superconducting fluxons  have been proposed  and realised in a series of flux cloning devices\cite{gulevich2006flux}, \cite{gulevich2006new}.  The  splitting of the Josephson fluxons has been produced by creating relatively large geometries (e.g. a T-shape) of the  long Josephson junction\cite{gulevich2006new}\cite{gulevich2017josephson}.  Using such flux-beam splitting it was possible to create a new device producing chaotic THz signals\cite{gulevich2019bridging}, therewith bridging the THz gap for chaotic sources. For a single electron the wave packet splitting may also lead to the formation of Majorana fermions,  which are topologicaly stable massless elementary excitations and therewith become building blocks of future quantum computers. To get a controllable, moveable Majorana fermion is a long-standing problem. Here it is demonstrated that graphene geometries provide unique conditions giving viability to the creation and control of Majorana fermions that are moving at the graphene Fermi (light equivalent) velocity. Such electron beam splitters provide a natural way to create quantum gates operating with Majorana fermions. 

\section{Conclusion}
The paper focuses on examining the behavior of electrons in Dirac matter, with the aim of improving quantum electron optics.  We found that the energy and shape of the excitation are crucial factors in minimizing the spread of the wavefunction when electrons propagate through nanoribbons. Additionally, the width of the nanoribbon channel is also an important factor in this process.

The work suggests that a balance between the topology (shape and arrangement of the nanoribbons), input energy, and pulse shape is necessary in designing efficient quantum electron optics. This type of system has been experimentally described for two-dimensional electron gases, and the work further demonstrated that splitting the wavefunction into nanoscale channels could lead to super-focusing of the electron\cite{Dubois2013}.  Besides graphene the Dirac spectrum does exist in a large variety of recently discovered new types of 2D materials such as covalently bonded  Stanene/Germanene with topological giant capacitance effects\cite{zhang2023covalent}.
Due to their large capacitance such materials may be favorable for electron optics.

We believe that this research has significant implications for the design of various quantum devices, such as electron microscopes\cite{hills2017current}, quantum computers, novel sensors\cite{villegas2020optical}, and quantum repeaters for signal enhancement. By understanding how electrons behave in Dirac matter and how to control their wavefunction propagation, researchers can design more efficient and precise quantum devices for various applications.

\section*{Acknowledgments}
DMF thanks QinetiQ for funding under an internal innovation and development grant,
FVK  is grateful to Khalifa University of Science and Technology under Award Nos FSU-2021-030/8474000371 and the EU H2020 RISE project TERASSE (H2020-823878).

\end{document}